\documentclass[onecolumn,secnumarabic,amssymb, nobibnotes, aps, prd]{revtex4}
\usepackage{amsmath} \usepackage{amssymb}
\usepackage{graphicx} 
\usepackage{epstopdf}
\usepackage{amsmath}
\usepackage{amsmath,amssymb,amsthm,amsfonts,mathrsfs,bm,verbatim}
\usepackage{graphicx,subfigure}

\newcommand{\bea}{\begin{eqnarray}}
\newcommand{\eea}{\end{eqnarray}}

\begin{document}

\title{Wormhole Stability of Charged Black Holes Under the Gravity of f(R)}%

\author{Wen-Xiang Chen$^{a}$}
\affiliation{Department of Astronomy, School of Physics and Materials Science, GuangZhou University, Guangzhou 510006, China}
\author{Yao-Guang Zheng}
\email{hesoyam12456@163.com}
\affiliation{Department of Physics, College of Sciences, Northeastern University, Shenyang 110819, China}


\begin{abstract}
In this paper, we propose that since Hawking radiation may be a kind of superradiation, the action of superradiation with preset bounds satisfies some higher-dimensional action structure, e.g., bosons in the preset springs of Kerr black hole. The superradiation meets the algebraic system of bosons on Kerr-Schild black holes. We think that the superradiation of bounded fermions to Kerr structure black holes benefits the mapping structure of wormholes. By analogy, we find the entropy algebraic expression of the wormhole in the charged black hole under f(R) gravity and then conduct thermodynamic geometric analysis to find out the stability condition of the wormhole.


\end{abstract}

\maketitle

\section{Introduction}
Wormhole, also known as the Einstein-Rosen Bridge, is a narrow tunnel that may exist in the universe connecting two different universes. Austrian physicist Ludwig Flemm first proposed the concept of wormholes in 1916. In the 1930s, Einstein and Nathan Rosen studied the gravitational field equations, assuming that black holes and white holes were connected through wormholes. The idea is that wormholes can be used to teleport or travel through time. Until now, scientists have observed no evidence of wormholes, which are thought to be difficult to distinguish from black holes.

To distinguish them from other types of wormholes, such as quantum wormholes in quantum states and in string theory, what is commonly known as a “Wormhole” should be called a “Spacetime hole.” Quantum wormholes in quantum states are generally called “Mini-wormholes.” There’s a big difference.

One of the properties of black holes is that they get what’s called a “Mirror Universe” on the other side. But Einstein didn’t take the solution seriously because we couldn’t get through. The bridge Einstein, which connects the two universes, was thought to be a mathematical trick.

But in 1963, a study by mathematicians at the Roy Kerr in New Zealand found that assuming that any collapsing star would rotate, it would become a dynamic black hole when it formed a black hole; Karl Schwarzschild’s static black hole is not the best physical explanation. However, a regular star will turn into a flat structure and not form a singularity. That is to say: the gravitational field is not infinite. This leads us to a startling conclusion: if we launch an object or spacecraft along the axis of rotation of a spinning black hole, it might, in principle, survive the center’s gravitational field and enter the mirror universe. In this way, Einstein was like a gateway between two regions of space and time, a “Wormhole.”

Wormholes are, in theory, hyperspace tunnels that link white holes to black holes, which are ubiquitous but fleeting. But some have hypothesized that a strange substance could keep the wormhole open. It has also been postulated that if there is a peculiar matter called Phantom matter, which has both negative energy and negative mass, it can create repulsive effects to prevent wormholes from closing. The strange material deflects light, signaling the discovery of a wormhole. But these theories have too many untested hypotheses to be convincing.

According to ER = EPR, two black holes can be entangled and then separated to form a wormhole between them (basically, a shortcut). Similarly, from the perspective of string theory, entanglement between two quarks would have the same effect.

These theoretical results provide support for some new theories. These new theories suggest that gravity and its physical properties are not fundamental but arise from quantum entanglement. Although quantum mechanics correctly describes interactions at the microscopic level, it has not yet been able to explain gravity. The theory of quantum gravity should be able to demonstrate that classical gravity is not fundamental, as proposed by Albert Einstein, but arises from a more fundamental quantum phenomenon.

The Schwinger effect creates pairs of entangled particles from a vacuum that can be trapped in an electric field, preventing them from annihilating back into the void. These trapped particles are entangled and can be mapped to four dimensions (a representation of spacetime). In contrast, physicists believe that gravity exists in a fifth dimension, bending and warping spacetime according to Einstein’s law.

According to the holographic principle, all events in the fifth dimension can be transformed into events in the other four sizes. Therefore, as entangled particles are generated, wormholes are also developed. More fundamentally, the argument suggests that gravity and its ability to bend spacetime come from quantum entanglement.

In 1972, Press and Teukolsky proposed that it is possible to create a black hole bomb by adding a mirror to the outside of the black hole (a scattering process that, according to current interpretations, involves classical and quantum mechanics)\cite{manogue1988klein,cardoso2004black,berti2006eigenvalues,herdeiro2014kerr,hartman2010holographic}).

When a  bosonic wave is impinging upon a rotating black hole, the wave reflected by the event horizon will be amplified if the wave frequency $ \omega$ lies in the following superradiant regime\cite{hod2012instability}
\begin{equation}\label{superRe}
   0<\omega < m\Omega_H ,{{\Omega }_{H}}=\frac{a}{r_{+}^{2}+{{a}^{2}}},
  \end{equation}
where  $m$ is an azimuthal number of the bosonic wave mode, $\Omega_H$ is the angular velocity of the black hole horizon. This amplification is called superradiant scattering. Therefore, the superradiation process can extract the rotational energy of the black hole. Suppose there is a mirror between the event horizon of the black hole and the infinite space. In that case, the amplified wave will scatter back and forth and grow exponentially, which will cause the superradiation of the black hole to become unstable.

In this paper, we propose that since Hawking radiation may be a kind of superradiation, the action of superradiation with preset bounds satisfies some higher-dimensional action structure, e.g., bosons in the preset springs of Kerr black hole. The superradiation meets the algebraic system of bosons on Kerr-Schild black holes. We think that the superradiation of bounded fermions to Kerr structure black holes benefits the mapping structure of wormholes. By analogy, we find the entropy algebraic expression of the wormhole in the charged black hole under f(R) gravity and then conduct thermodynamic geometric analysis to find out the stability condition of the wormhole.

\section{{Fermionic scattering}}
Now\cite{brito2015superradiance} let us consider the Dirac equation for a spin-$\frac{1}{2}$ massless fermion $\Psi$, minimally coupled to tand he same EM potential $A_{\mu}$ as in Eq..
\begin{equation}
\gamma^{\mu}\Psi_{;\mu}=0\,,
\end{equation}
where $\gamma^{\mu}$ are the four Dirac matrices satisfying the anticommutation relation $\{\gamma^{\mu},\gamma^{\nu}\}=2g^{\mu\nu}$. 
The solution to takes the form $\Psi=e^{-i\omega t}\chi(x)$, where $\chi$ is a two-spinor given by
\begin{equation}
\chi=\begin{pmatrix}f_1(x)\\f_2(x)\end{pmatrix}\,.
\end{equation}
Using the representation
\begin{equation}
\gamma^0=\begin{pmatrix}i&0\\0&-i\end{pmatrix}\,,\,\,\gamma^1=\begin{pmatrix}0&i\\-i&0\end{pmatrix}\,,
\end{equation}
the functions $f_1$ and $f_2$ satisfy the system of equations:
\begin{equation}
df_1/dx-i(\omega-eA_0)f_2=0,
df_2/dx-i(\omega-eA_0)f_1=0.
\end{equation}
One set of solutions can be once more formed by the `in' modes, representing a flux of particles coming from $x\to-\infty$ being partially reflected (with reflection amplitude $|\mathcal{R}|^2$) and partly transmitted at the barrier
\begin{equation}
\left(f_1^{\rm in},f_2^{\rm in}\right)=                                               \left(\mathcal{I}e^{i\omega x}-\mathcal{R} e^{-i\omega x},\mathcal{I} e^{i\omega x}+\mathcal{R} e^{-i\omega x}\right) \quad \text{as} \,\, x\to-\infty
\end{equation}
   \begin{equation}                                            \left(\mathcal{T}e^{ikx},\mathcal{T}e^{ikx}\right) \quad \hspace{2.9cm} \text{as}\,\, x\to+\infty
\end{equation}

On the other hand, the conserved current associated with the Dirac equation is given by
$j^{\mu}=-e\Psi^{\dagger}\gamma^{0}\gamma^{\mu}\Psi$ and, by equating the latter at $x\to-\infty$ and $x\to+\infty$, we find some general relations between the reflection and the transmission coefficients, in particular,
\begin{equation}
\left|\mathcal{R}\right|^2=|\mathcal{I}|^2-\left|\mathcal{T}\right|^2.
\end{equation}
Therefore, $\left|\mathcal{R}\right|^2\leq |\mathcal{I}|^2$ for any frequency, showing no superradiance for fermions. The same kind of relationship can be found for massive fields. 

The reflection and transmission coefficients depend on the specific shape of the potential $A_0$. However, one can easily show that the Wronskian
\begin{equation}
W=\tilde{f}_1 \frac{d\tilde{f}_2}{dx}-\tilde{f}_2\frac{d\tilde{f}_1}{dx}\,,
\end{equation}
between two independent solutions, $\tilde{f}_1$ and $\tilde{f}_2$, of is conserved.
From the equation, on the other hand, if $f$ is a solution, then its complex conjugate $f^*$ is another linearly independent solution. We find$\left|\mathcal{R}\right|^2=|\mathcal{I}|^2-\frac{\omega-eV}{\omega}\left|\mathcal{T}\right|^2$.Thus,for
$0<\omega<e V$, it is possible to have superradiant amplification of the reflected current, i.e., $\left|\mathcal{R}\right|>|\mathcal{I}|$. 
Other potentials can be resolved entirely, which can also show superradiation explicitly.

The difference between fermions and bosons comes from the intrinsic properties of these two kinds of particles. Fermions 
have positive definite current densities and bounded transmission amplitudes $0\leq \left|\mathcal{T}\right|^2\leq 
|\mathcal{I}|^2$, while for bosons, the current density can change its sign as it is partially transmitted and the 
transmission amplitude can be negative, $-\infty < \frac{\omega-eV}{\omega}\left|\mathcal{T}\right|^2\leq 
|\mathcal{I}|^2$. From the point of view of quantum field theory, due to strong electromagnetic fields, one can understand this process as a spontaneous pair generation phenomenon (see, for example). The number of spontaneously produced iron ion pairs in a given state is limited by Poly's exclusion principle, while bosons do not have this limitation.

We can pre-set the boundary conditions $e{A_0(x)} = -{y}{\omega}$(which can be ${\mu} = {-y}{\omega}$)\cite{chen2020possibility}\cite{chen2019superradiant}.

\section{{ The relation between the algebraic structure of bosons on Kerr-Schild black holes and wormhole}}
 We will prove that the Hawking radiation of the Kerr black hole can be understood as the flux that offsets the gravitational anomaly. The key is that near the horizon, the scalar field theory in the spacetime of a 4-dimensional Kerr black hole can be simplified to a 2-dimensional field theory. Since spacetime is not spherically symmetric, this is an unexpected result.
 
In Boyer-Linquist coordinates, Kerr metric reads\cite{Keiju}
\begin{equation}
\begin{split}
d s^{2}=-\frac{\Delta-a^{2} \sin ^{2} \theta}{\Sigma} d t^{2}-2 a \sin ^{2} \theta \frac{r^{2}+a^{2}-\Delta}{\Sigma} d t d \phi \\
+\frac{\left(r^{2}+a^{2}\right)^{2}-\Delta a^{2} \sin ^{2} \theta}{\Sigma} \sin ^{2} \theta d \phi^{2}+\frac{\Sigma}{\Delta} d r^{2}+\Sigma d \theta^{2}
\end{split}
\end{equation}
\begin{equation}
\Sigma=r^{2}+a^{2} \cos ^{2} \theta,
\Delta=r^{2}-2 M r+a^{2} 
=\left(r-r_{+}\right)\left(r-r_{-}\right).
\end{equation}

The action for the scalar field in the Kerr spacetime is
\begin{equation}
\begin{aligned}
S[\varphi]=& \frac{1}{2} \int d^{4} x \sqrt{-g} \varphi \nabla^{2} \varphi \\
=& \frac{1}{2} \int d^{4} x \sqrt{-g} \varphi \frac{1}{\Sigma}\left[-\left(\frac{\left(r^{2}+a^{2}\right)^{2}}{\Delta}-a^{2} \sin ^{2} \theta\right) \partial_{t}^{2}\right.\\
&-\frac{2 a\left(r^{2}+a^{2}-\Delta\right)}{\Delta} \partial_{t} \partial_{\phi}+\left(\frac{1}{\sin ^{2} \theta}-\frac{a^{2}}{\Delta}\right) \partial_{\phi}^{2} \\
&\left.+\partial_{r} \Delta \partial_{r}+\frac{1}{\sin \theta} \partial_{\theta} \sin \theta \partial_{\theta}\right] \varphi
\end{aligned}
\end{equation}
Taking the limit r → r+ and leaving the dominant terms,
we have
\begin{equation}
\begin{aligned}
S[\varphi]=& \frac{1}{2} \int d^{4} x \sin \theta \varphi \left[-\frac{\left(r_{+}^{2}+a^{2}\right)^{2}}{\Delta} \partial_{t}^{2}\right.\\
&\left.-\frac{2 a\left(r_{+}^{2}+a^{2}\right)}{\Delta} \partial_{t} \partial_{\phi}-\frac{a^{2}}{\Delta} \partial_{\phi}^{2}+\partial_{r} \Delta \partial_{r}\right] \varphi
\end{aligned}
\end{equation}
Now we transform the coordinates to the locally non-rotating coordinate system by
\begin{equation}
\left\{\begin{array}{l}
\psi=\phi-\Omega_{H} t \\
\xi=t
\end{array}\right.
\end{equation}
where
\begin{equation}
\Omega_{H} \equiv \frac{a}{r_{+}^{2}+a^{2}}.
\end{equation}
 We can rewrite the action
 \begin{equation}
S[\varphi]=\frac{a}{2 \Omega_{H}} \int d^{4} x \sin \theta \varphi\left(-\frac{1}{f(r)} \partial_{\xi}^{2}+\partial_{r} f(r) \partial_{r}\right) \varphi
\end{equation}
When $\sin \theta$ = 0, the pull equation for action can conform to the above form, but the boundary becomes 0. However, if the boundary conditions are preset, the boundary conditions${\mu} = {-y}{\omega}$(y takes a larger number) act as $\sin \theta$.The effective action form of Fermion superradiation satisfies the effective action form of the wormhole and is not necessarily at the boundary of the event horizon. When the boundary condition is preset, a new path is obtained
\begin{equation}
S[y,\varphi]=\frac{a}{2 \Omega_{H}} \int d^{4} x (-y)e^{y}\varphi \left(-\frac{1}{f(r)} \partial_{\xi}^{2}+\partial_{r} f(r) \partial_{r}\right) \varphi.
\end{equation}

\section{{Wormhole Stability of Charged Black Holes Under the Gravity of f(R)}}
In this paper, action is linked to entropy\cite{Caravelli}. From the analogy in the previous chapter, we get an algebra expression of the action of the wormhole. Through the thermodynamic geometric analysis of its entropy, we get the stability analysis of the wormhole. We consider the spherically symmetric solution of $f(R)$ gravity with constant curvature $R_{0}$ (or initial Ricci curvature) under the model, such as the RN black hole solution. $\left(R_{0}=0\right)$, where $g(r)$ is in the form of $f(r )=1-\frac{2 M}{r}-\frac{R_{0} r^{2}}{12}+\frac{Q^2}{r^2}$, its  metric is as follows \cite{Caravelli}
\begin{equation}
\begin{aligned}
    \mathrm{d}s^{2} =-\left(1-\frac{2 M}{r}-\frac{R_{0} r^{2}}{12}+\frac{Q^2 }{r^2}\right) \mathrm{d} t^{2} 
    +\left(1-\frac{2 M}{r}-\frac{R_{0} r^{2}}{12}+\frac{Q^2}{r^2}\right)^{- 1} \mathrm{~d} r^{2}
    +r^{2} d \theta^2+r^{2} \sin^{2}{\theta}d\varphi^2,
\end{aligned}
\end{equation}
where $R_0$ is the cosmological constant.

This section summarizes the expression of the black hole correction thermodynamic entropy caused by small fluctuations near the equilibrium\cite{Chen}. For this, let us first define the density of states with fixed energy as(with the natural unit, $G=\hbar=c=1)$
\begin{equation}
\rho(E)=\frac{1}{2 \pi i} \int_{c-i \infty}^{c+i \infty} e^{\mathcal{S}(\beta)} d \beta
\end{equation}
The exact entropy $\mathcal{S}(\beta)=\log Z(\beta)+\beta E$ clearly depends on the temperature $T\left(=\beta^{-1}\right)$. So, this (actual entropy) is not just its equilibrium value. The exact entropy corresponds to the sum of the entropy of the thermodynamic system subsystems. The thermodynamic system is small enough to be considered in equilibrium. To study the form of exact entropy, we solve the complex integral by considering the method of steepest descent around the saddle point $\beta_0\left(=T_H^{-1}\right)$ so that $\left.\frac{\partial \mathcal{S}(\beta)}{\partial \beta}\right|_{\beta=\beta_0}=0$. Now, the Taylor expansion of the exact entropy is performed around the saddle point beta $=\beta_0$ causes
\begin{equation}
\mathcal{S}(\beta)=\mathcal{S}_0+\frac{1}{2}\left(\beta-\beta_0\right)^2\left(\frac{\partial^2 \mathcal{S}(\beta)}{\partial \beta^2}\right)_{\beta=\beta_0}+(\text { higher order terms })
\end{equation}
and
\begin{equation}
\rho(E)=\frac{e^{\mathcal{S}_0}}{2 \pi i} \int_{c-i \infty}^{c+i \infty} \exp \left[\frac{1}{2}\left(\beta-\beta_0\right)^2\left(\frac{\partial^2 \mathcal{S}(\beta)}{\partial \beta^2}\right)_{\beta=\beta_0}\right] d \beta
\end{equation}
We get that
\begin{equation}
\rho(E)=\frac{e^{\mathcal{S}_0}}{\sqrt{2 \pi\left(\frac{\partial^2 \mathcal{S}(\beta)}{\partial \beta^2}\right)_{\beta=\beta_0}}}
\end{equation}
where $c=\beta_0$ and $\left.\frac{\partial^2 \mathcal{S}(\beta)}{\partial \beta^2}\right|_{\beta=\beta_0}>0$ are chosen.

The entropy
\begin{equation}
S(\mathrm{L})=(-y) e^y \varphi\left(-\frac{1}{f(r)} \partial_{\xi}^2+\partial_r f(r) \partial_r\right) \varphi.
\end{equation} The metric of Ruppeiner geometry is
\begin{equation}
d s^2=-\frac{\partial^2 S\left(M, Q\right)}{\partial X^\alpha \partial X^\beta} \Delta X^\alpha \Delta X^\beta.
\end{equation}
\begin{equation}
g_{MM}=\frac{8\varphi^2 e^y \varphi y}{r^2\left(1+\frac{Q^2}{r^2}-\frac{2 M}{r}-\frac{r^2 {R_0}}{12}\right)^3}
\end{equation}
\begin{equation}
g_{QQ}=\frac{4 e^y \varphi\left(\varphi+\frac{72\varphi^2 r^5\left(36 Q^2+r\left(24 M-12 r+r^3 {R_0}\right)\right)}{\left(12 Q^2-r\left(24 M-12 r+r^3 {R_0}\right)\right)^3}\right) y}{r^3}
\end{equation}
\begin{equation}
g_{MQ}=-\frac{8 \varphi^2 e^y \varphi Q y}{\left(-2 M+\frac{Q^2}{r}+r-\frac{r^3 {R_0}}{12}\right)^3}.
\end{equation}The curvature scalar is
\begin{equation}
R(S)=\frac{9 e^{-y} r^3\left(12 Q^2-r\left(24 M-12 r+r^3 {R_0}\right)\right)\left(-24 \varphi^2 r^5+\operatorname{\varphi}\left(-12 Q^2+r\left(24 M-12 r+r^3 {R_0}\right)\right)^2\right)}{\varphi\left(-72 \varphi^2 r^5+\operatorname{\varphi}\left(-12 Q^2+r\left(24 M-12 r+r^3 {R_0}\right)\right)^2\right)^2 y}.
\end{equation}

\section{{Summary}}
In this paper, we propose that since Hawking radiation may be a kind of superradiation, superradiation with a preset boundary satisfies some higher-dimensional interaction structure, for example, on the boson in the preset limit of the Kerr black hole. Superradiation benefits the algebraic system of bosons on Kerr-Schild black holes. We think that the superradiation of bounded fermions to Kerr structure black holes meets the mapping structure of wormholes. By analogy, the algebraic expression of the entropy of the wormhole in the charged black hole under the gravitational action of f(R) is obtained. Then the thermodynamic geometric analysis is performed to find the stable condition of the wormhole. The wormhole is stable when R(S) is not equal to infinity.

\end{document}